\documentclass[a4paper]{article}
\usepackage{graphicx}
\usepackage{INTERSPEECH2020}
\usepackage{makecell}
\usepackage{float}
\usepackage{hyperref}
\usepackage{cite}
\usepackage[labelfont=bf]{caption}
\usepackage{amsmath}

\title{Audio Dequantization for High Fidelity Audio Generation in Flow-based Neural Vocoder}
\name{Hyun-Wook Yoon$^1$, Sang-Hoon Lee$^2$, Hyeong-Rae Noh$^2$, Seong-Whan Lee$^{2,3}$\thanks{This work was supported by Institute of Information \& communications Technology Planning \& Evaluation (IITP) grant funded by the Korea government (MSIT) (No. 2019-0-00079, Department of Artificial Intelligence, Korea University), the Magellan Division of Netmarble Corporation, and the Seoul R\&BD Program(CY190019).}}

\address{
   $^1$Department of Computer and Radio Communications Engineering, Korea University, Seoul, Korea\\
  $^2$Department of Brain and Cognitive Engineering, Korea University, Seoul, Korea\\
  $^3$Department of Artificial Intelligence, Korea University, Seoul, Korea}

\email{\{hw\_yoon, sh\_lee, hr\_noh, sw.lee\}@korea.ac.kr}

\begin{document}

\maketitle
\begin{abstract}
    In recent works, a flow-based neural vocoder has shown significant improvement in real-time speech generation task. The sequence of invertible flow operations allows the model to convert samples from simple distribution to audio samples. However, training a continuous density model on discrete audio data can degrade model performance due to the topological difference between latent and actual distribution. To resolve this problem, we propose audio dequantization methods in flow-based neural vocoder for high fidelity audio generation. Data dequantization is a well-known method in image generation but has not yet been studied in the audio domain. For this reason, we implement various audio dequantization methods in flow-based neural vocoder and investigate the effect on the generated audio. We conduct various objective performance assessments and subjective evaluation to show that audio dequantization can improve audio generation quality. From our experiments, using audio dequantization produces waveform audio with better harmonic structure and fewer digital artifacts. 
\end{abstract}

\section{Introduction}
Most speech synthesis models take two-stage procedures to generate waveform audio from the text. First stage generates spectrogram conditioned on linguistic features such as text or phoneme.\cite{jia2018transfer:tts,wang2017tacotron:tts,park2019phonemic:1.1:tts,taigman2017voiceloop:1.2:tts,shen2018natural:tts} In second stage, generally refer to as vocoder stage, audio samples are generated through model capable of estimating audio samples from the acoustic features. Traditional approaches estimated audio samples either directly from the spectral density model\cite{griffin1984signal:3} or hand-crafted acoustic model\cite{kawahara2006straight:4,morise2016world:5}, but these approaches tended to produce low-quality audio.

After the emergence of the WaveNet\cite{oord2016wavenet:vocoder}, models that generate audio samples on previously generated samples had shown exceptional works in the field.\cite{tamamori2017speaker:vocoder,hayashi2017investigation:vocoder,arik2017deep:ttsvocoder}. Nevertheless, dilated causal convolution networks used in the model require sequential generation process during the inference, which infers that real-time speech synthesis is hard to achieve because parallel inference can't be utilized. For this reason, generating high-quality waveform audio in real-time has become a challenging task.

To overcome the structural limitation of the auto-regressive model, most of the recent works are focused on non-autoregressive models such as knowledge distillation\cite{ping2018clarinet:10,oord2017parallel:11}, generative adversarial network\cite{engel2019gansynth:gan,neekhara2019expediting:gan,yamamoto2019probability:gan,kumar2019melgan:12,yamamoto2020parallel}, and flow-based generative model\cite{prenger2019waveglow:6,kim2018flowavenet:7}. We focus on the flow-based generative model since it can model highly flexible approximate posterior distribution in variational inference\cite{rezende2015variational:normalizingflow}. The transformation from a single data-point to a Gaussian noise is one-to-one, which makes the parallel generation possible. However, we have to acknowledge that audio samples are discrete data. In other words, naive modeling of a continuous probability density on discrete data can produce arbitrary high likelihood on discrete location\cite{theis2015note:9,hoogeboom2020learning:17}. This can lead to degraded generation performance in flow-based neural vocoder. Therefore, \textit{dequantization} is required before the transformation.

In this paper, we present various audio dequantization schemes that can be implemented in the flow-based neural vocoder. In image generation, adding continuous noise to data-points to \textit{dequantize} the data is commonly used. However, to the best of our knowledge, the effectiveness of data dequantization in audio domain is still an unknown area, so further investigation is needed. Unlike pixels of the image, audio samples are bounded to signed integer. To overcome this domain issue, we either normalize range of noise values or range of audio samples with different normalization method. In addition, we adapt flow block from flow-based neural vocoder to generate more flexible noises known as \textit{variational dequantization}\cite{ho2019flow++:15}.

\begin{figure}[htb]
    \begin{minipage}[b]{1.0\linewidth}
        \centering
        \includegraphics[width=\linewidth]{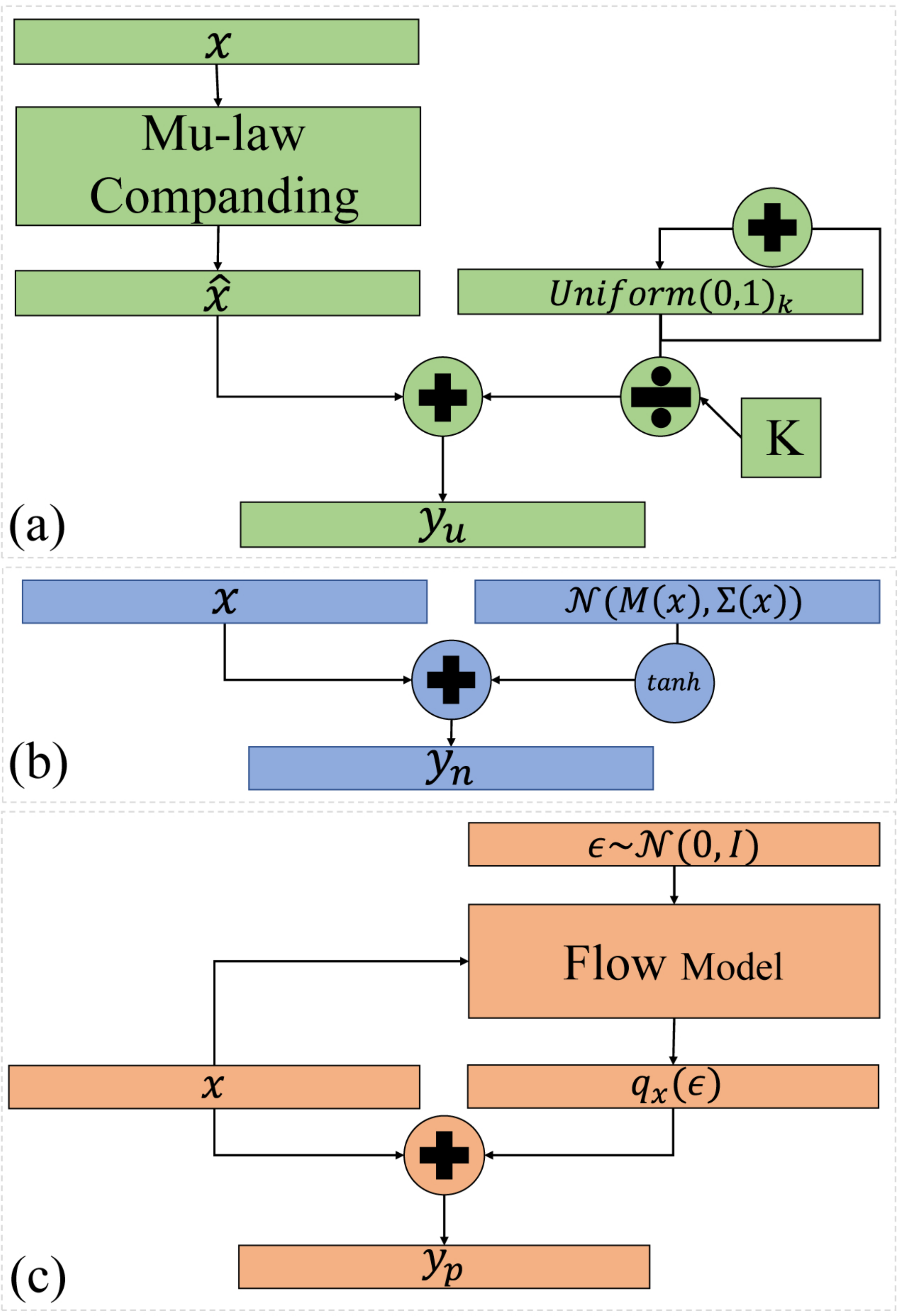}
    \end{minipage}
    \caption{Examples of audio dequantization for flow-based neural vocoder. (a) represents uniform dequantization. (b) represents Gaussian dequantization. (c) represents variational dequantization.}
\end{figure}

\section{Flow-based Neural Vocoder}
FloWaveNet\cite{kim2018flowavenet:7} and WaveGlow\cite{prenger2019waveglow:6} are two pioneers in flow-based neural vocoders. Both models are based on normalizing flow\cite{rezende2015variational:normalizingflow}. Two main contributions that they share are training simplicity and faster generation. Since they use a single invertible flow network repeatedly, model structure is intuitive. Moreover, optimization can be easily done with a single log-likelihood loss function. During inference, random noises of equal length to the product of frames of mel-spectrogram and hop-size are sampled from the spherical Gaussian distribution and simultaneously converted to audio samples. As a result, flow-based neural vocoders can produce waveform signal as fast as other non-autoregressive models.

In general, flow-based neural vocoder requires three steps: squeeze, flow, and shuffle. In the squeezing step, the temporal dimension of the feature is reduced whereas channels of the feature are increased. According to FloWaveNet\cite{kim2018flowavenet:7}, this operation increase the size of the receptive field like dilated convolutions layer from the WaveNet\cite{oord2016wavenet:vocoder}. During the flow step, multiple blocks of flow operate affine transformation on the half of input vectors. In detail, half input vectors are used to predict shift and scale parameters for the other half in each flow operation. Lastly, the shuffle step mixes elements of data, giving flexibility in transformation.

Although both models have similar concepts, they use different techniques in detail. FloWaveNet\cite{kim2018flowavenet:7} defines one squeeze operation and multiple flow operations as a single context block and duplicates this context block to form the flow process. The model also implements the activation normalization layer suggested in Glow\cite{kingma2018glow:8} before the coupling layer to stabilize training. After each flow, the model simply swap odd and even elements of vectors to shuffle features. WaveGlow\cite{prenger2019waveglow:6} operates the squeezing step only once during the process. Also, the model adapts invertible 1x1 convolution from Glow to operate the shuffle step before each flow operation.

\section{Audio Dequantization}
A raw audio is stored in computer digitally. In other words, values of the audio are formed as discrete representations. Therefore, naively transforming audio samples into Gaussian noise can lead to arbitrary high likelihood on values of data in flow-based neural vocoder. To resolve this issue, we adapt the idea of adding noise to each of the data-point to \textit{dequantize} discrete distribution data in image generation task\cite{theis2015note:9}. In image, a pixel \(x\) is represented as a single discrete value in \(\{ 0, 1, \dots, 255\}\), so dequantized data \(y\) can be formulated as \(y=x+u\), where \(u\) represents D components of noise bounded to \([0, 1)^{D}\). 

Unlike image, raw audio encoded in 16-bit WAV contains 15-bit of negative and positive integer values, which can be represented as \(\{-32,767, \dots, 32,767\}\). To apply data dequantization to raw audio, either range of audio samples must be compressed to 8-bit unsigned integer, or the range of dequantized data has to be within the range of \((-1, 1)^{D}\). For the proper audio dequantization, we present three different methods in the following sections.

\subsection{Uniform Dequantization}
In \cite{theis2015note:9}, authors note that optimizing the continuous model \(p_{model}(y)\) on the dequantized data \(y\sim p_{data}\) can closely optimize the discrete model \(P_{model}(x)\) on the origianl data \(x\sim P_{data}\) through Jensen's inequality, which can be formulated as below:
\begin{align}
  &\int p_{data}(y)\log p_{model}(y) dy\\
  &= \sum_{x}P_{data}(x)\int_{[0,1)^{D}}\log p_{model}(x+u)du\\
  &\leq \sum_{x} P_{data}(x)\log \int_{[0,1)^{D}} p_{model}(x+u)du\\
  &= \mathbb{E}_{x\sim P_{data}} \left[\log P_{model}(x) \right]
\end{align}

Since the uniform noise is bounded to \([0,1)^{D}\), values of audio samples have to be bounded to unsigned integer. For this purpose, we preprocess raw audio with nonlinear companding method called '\textit{mu-law companding}'\cite{yoshimura2018mel}. This method can significantly reduce the range of audio samples while minimizing the quantization error. Redistribution equation of the method can be expressed as:
\begin{align}
    \hat{x}=sign(x)(\frac{ln(1+\mu|x|)}{ln(1+\mu)})
\end{align}
where \(sign\) function represents sign of value x, and \(\mu\) represents integers that x values are mapped. We set \(\mu\) to \(255\) to apply 8-bit \textit{mu-law companding}. Then, we add random noise from uniformly distributed function formulated as \(Unif(0,1)\).

We assume that companding audio with lossy compression can possibly produce noisy output. Therefore, we implement \textit{iw(importance-weighted) dequantization} proposed in \cite{hoogeboom2020learning:17} to improve generation quality. In the paper, authors demonstrate that sampling noise multiple times can directly approximate the objective log-likelihood, which can lead to better log-likelihood performance. As a result, we define uniformly dequantized data \(y_{u}\) as:
\begin{align}
    y_{u} = \hat{x} + \frac{1}{K}\sum_{k=1}^{K} Unif(0,1)_{k}
\end{align}
where \(Unif(0,1)\) defines noise sampled uniformly from \([0, 1)^{D}\). We set K to 10. 

To compare the performance of model depending on \textit{iw dequantization}, we refer to uniform dequantization with \textit{iw dequantization} as \textit{Uniform\_IW} and only uniform deqauntization model as \textit{Uniform}.

\subsection{Gaussian Dequantization}
In flow-based neural vocoder, discrete data distribution is transformed into a spherical Gaussian distribution. In other words, dequantizing data distribution to normal distribution can be more optimal choice. With this though in mind, we formulate Gaussian dequantization motivated from logistic-normal distributions \cite{atchison1980}. Random noise samples are generated from normal distribution formulated as \(\mathcal{N}(\mu, \sigma^{2})\), where mean and variance are calculated from the given data batch. To properly implement in audio domain, we apply a hyperbolic tangent function to normalize noise boundary at \((-1, 1)^{D}\). As a result, normally dequantized data \(y_{n}\) can be formulated as:
\begin{equation}
\begin{split}
     y_{n}=x+\tanh(\mathcal{N}(M(x_{b})), \Sigma(x_{b}))\\
\end{split}
\end{equation}
where \(M(x_{b})\) and \(\Sigma(x_{b})\) represent mean and variance of batch group \(x_{b}\). 

To compare model performance between conventional and improved method, we refer to the conventional method suggested in \cite{atchison1980} as \textit{Gaussian\_Sig} and proposed method as \textit{Gaussian\_Tanh}.

\subsection{Variational Dequantization}
Instead of adding noise from known distribution, noise distribution can be formulated through a neural network such as a flow-based network. Flow++ \cite{ho2019flow++:15} suggests that if the noise samples \(u\) are generated from conditional probability model \(q(u|x)\), probability distribution of original data can be estimated as follows:
\begin{align}
    P_{model}(x):=\int_{[0,1)^{D}}q(u|x)\frac{p_{model}(x+u)}{q(u|x)}du
\end{align}

Then, we can obtain the variational lower-bound on the log-likelihood function by applying Jensen's inequality as below:

\begin{align}
  &\mathbb{E}_{x\sim P_{data}} \left[\log P_{model}(x) \right]\\
  &=\mathbb{E}_{x\sim P_{data}} \left[\log \int_{[0,1)^{D}} q(u|x)\frac{p_{model}(x + u)}{q(u|x)}du \right]\\
  &\geq \mathbb{E}_{x\sim P_{data}} \left[\int_{[0,1)^{D}} q(u|x)\log \frac{p_{model}(x + u)}{q(u|x)}du \right]\\
  &=\mathbb{E}_{x\sim P_{data}} \mathbb{E}_{u\sim q(u|x)} \left[\log\frac{p_{model}(x + u)}{q(u|x)}\right]
\end{align}

As a result, dequantized data \(y_{p}\) from variational dequantization can be defined as:
\begin{align}
    y_{p}=x+q_{x}(\epsilon)
\end{align}
where \(\epsilon \sim p(\epsilon)=\mathcal{N}(\epsilon;0,I)\).

To implement variational dequantization in flow-based neural vocoder, we modify flow model from FloWaveNet\cite{kim2018flowavenet:7}. We set initial input as 1-dimensional noise vector generated from spherical Gaussian distribution \(\mathcal{N}(\epsilon;0,I)\), where the length is equal to target audio. In each context block, single squeeze step and multiple flow steps are operated. In each flow step, an affine transformation conditioned on target audio is applied to the half of squeezed vector. At the end, the vector is flattened, and hyperbolic tangent function is applied to fit range of audio domain. Negative log likelihood from the dequantizer is trained jointly with the flow-based neural vocoder. 

We set a total of 16 flow stacks as \textit{Flow\_Shallow} and 48 flow stacks as \textit{Flow\_Dense} to examine whether the dept of dequantization model is critical to the model performance.

\begin{figure*}
    \centering
    \includegraphics[width=\textwidth, height=180px]{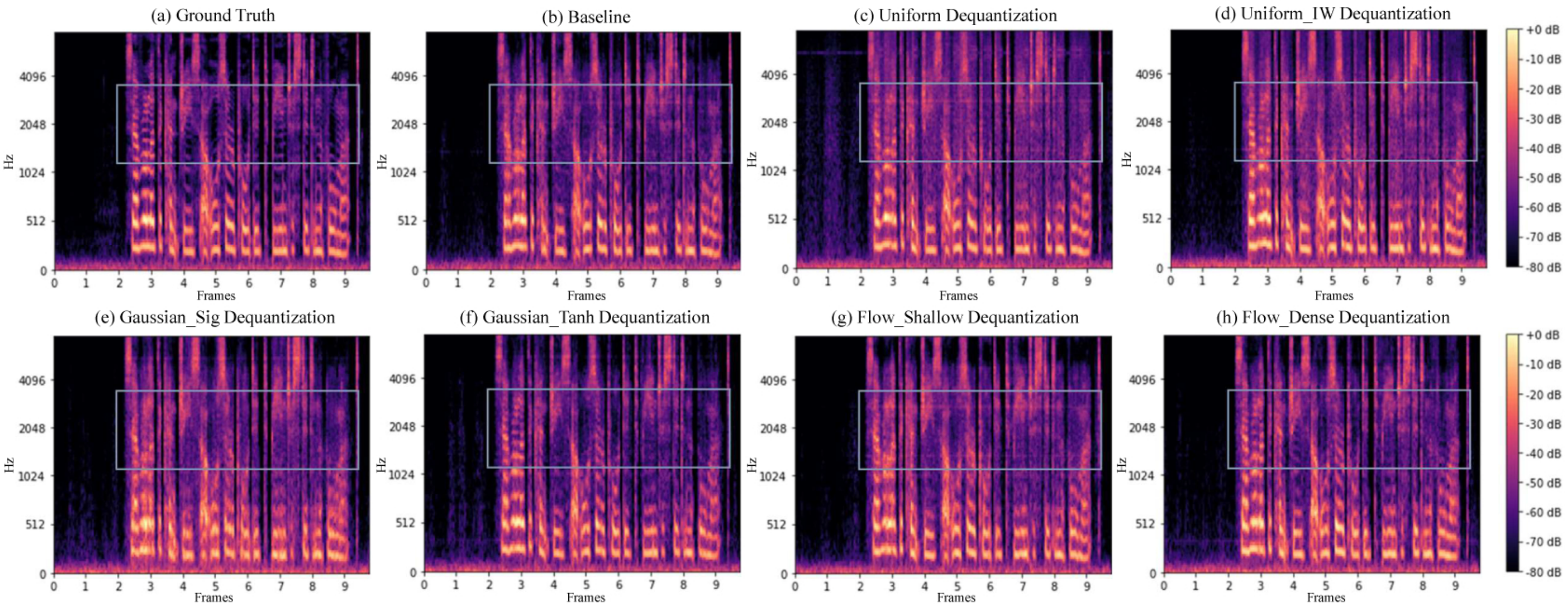}
    \caption{Mel-spectrogram converted from audio samples generated by baseline\cite{kim2018flowavenet:7} and proposed dequantization models. The blue bounding-box indicates the area where periodic noise appears and harmonic frequencies are presented.}
    \label{fig:mel-spec}
\end{figure*}

\section{Experimental Results and Analysis}
\subsection{Experimental Settings}

We set FloWaveNet\cite{kim2018flowavenet:7} as our baseline model and trained baseline with 6 different dequantization methods. Each model was trained with VCTK-Corpus\cite{veaux2016superseded} containing 109 native speakers English dataset. Since FloWaveNet and WaveGlow\cite{prenger2019waveglow:6} evaluated with only a single speaker dataset, we expanded the experiment on the model to a multimodal case where audio generation is much harder due to the larger variation among different speakers. In the dataset, we withdrew some corrupted audio files and used 44,070 audio clips. For each speaker, 70\% of data were used as training data, 20\% as validation data, and rest of them as test data. All clips were down-sampled from 48,000Hz to 22,050Hz. From each audio clip, 16,000 chunks were randomly extracted.

All models were trained on 4 Nvidia Titan Xp GPUs with a batch size of 8. We used Adam optimizer with a step size of \(1\times 10^{-3}\) and set the learning rate decay in every 200K iterations with a factor of 0.5. We trained each model for 600K iterations. 

\subsection{Subjective Evaluation}
For subjective evaluation, we conducted a subjective 5 scale MOS test on Amazon Mechanical Turk\footnote{\url{https://www.mturk.com/}}. Each participant was suggested to wear either earbuds or headphones for the eligible testing. Then they had to listen to 5 audio clips at least twice and rated naturalness of audio on a scale of 1 to 5 with 0.5 point increments. We explicitly instructed the participants to focus on the quality of audio. We collected approximately 3,000 samples for the evaluation. In the Table \ref{table:MOS}, models implemented audio dequantization show higher MOS than the baseline model which show that audio dequantization can improve audio quality. Except for the real audio, variational dequantization with a deeper layer receives the highest MOS. This shows that injecting noise with more complex distribution can produce more natural audio.

\begin{table}
\centering
\caption{Mean opinion score (MOS) results with 95\% confidence intervals on 150 randomly selected sentences in test set.}\label{table:MOS}
\begin{tabular}{lll}
\Xhline{3\arrayrulewidth}
\textbf{Methods}                       & \textbf{MOS} & \textbf{95\% CI}\\ \hline
Ground Truth                  & 4.489 & $\pm$0.016        \\
Baseline\cite{kim2018flowavenet:7}  & 2.898 & $\pm$0.043   \\
\textit{Uniform\_IW}   & 3.054 & $\pm$0.041         \\
\textit{Gaussian\_Tanh}  & 3.029 & $\pm$0.041         \\
\textit{Flow\_Dense}   & 3.267 & $\pm$0.036         \\ \Xhline{3\arrayrulewidth}
\end{tabular}
\end{table}

\subsection{Objective Evaluation}
Audio generated from the baseline model tended to have digital artifacts such as reverberation, trembling sound, and periodic noise. We assumed that the occurrence of these artifacts was due to the unnatural collapsing from the continuous density model on discrete data-points. To prove that audio dequantization can remove such artifacts, we conducted several quantitative evaluations in signal processing to compare the quality of audio. First we randomly selected 400 sentences in test set. Then, we conducted mel-cepstral distortion (MCD)\cite{407206:16}, global signal-to-noise ratio (GSNR)\cite{vondrasek2005methods:18}, segmental signal-to-noise ratio (SSNR)\cite{hayashi2017investigation:vocoder}, and root mean square error of fundamental frequency (RMSE\(_{f0}\))\cite{hayashi2017investigation:vocoder}. All equations for the evaluation can be calculated as follows:
\begin{align}
    &MCD_{13}[dB]=\frac{1}{T}\sum_{t=0}^{T-1}{\sqrt{\sum_{k=1}^{K}{(c_{t,k}-c_{t,k}^{'})^{2}}}}\\
    &GSNR[dB]=10log\frac{\sigma_{s}^{2}}{\sigma_{r}^{2}}\\    
    &SSNR[dB]=10log_{10}(\frac{\sum_{n=0}^{M} x_{s}(n)^{2}}{\sum_{n=0}^{M}(x_{s}(n)-y_{r}(n))^{2}})\\    
    &RMSE_{f0}[cent]=1200\sqrt{(log_{2}(F_{r})-log_{2}(F_{s}))^{2}}
\end{align}
where \(c_{t,k}\), \(c_{t,k}^{'}\) represent original and synthesized k-th mel frequency cepstral coefficient (MFCC) of t-th frame, \(\sigma_{s}^{2}\), \(\sigma_{r}^{2}\) represent power of speech signal and noise, \(x_{s}(n)\), \(y_{r}(n)\) represent raw and synthesized waveform sample at time n, and \(F_{r}\), \(F_{s}\) represent fundamental frequency of raw and synthesized waveform.

\begin{table}
\centering
\caption{MCD (dB) results with 95\% confidence intervals.}\label{table:MCD}
\begin{tabular}{lll}
    \Xhline{3\arrayrulewidth}
    $\textbf{Methods}$ & $\textbf{MCD}_{13}$ & $\textbf{95\% CI}$ \\ \hline
    Baseline\cite{kim2018flowavenet:7}   & 3.455 & $\pm$0.032 \\ \hline
    \textit{Uniform}    & 4.139 & $\pm$0.041 \\
    \textit{Uniform\_IW}    & 3.567 & $\pm$0.032 \\ \hline
    \textit{Gaussian\_Sig}  & 3.739 & $\pm$0.039 \\
    \textit{Gaussian\_Tanh} & 3.355 & $\pm$0.031 \\ \hline
    \textit{Flow\_Shallow} & 3.484 & $\pm$0.032 \\
    \textit{Flow\_Dense}   & 3.401 & $\pm$0.032 \\ \Xhline{3\arrayrulewidth}
\end{tabular}
\end{table}

In MCD, we compared all models including \textit{Uniform} and \textit{Gaussian\_Sig} to see the improvement within the modification. In Table \ref{table:MCD}, dequantizations with modified methods show better performance than the conventional methods. \textit{Gaussian\_Tanh} and \textit{Flow\_Dense} score relatively lower MCD than baseline, which shows that both models can produce better audio quality than the baseline model. There was no significant performance difference between the two variational dequantization methods.  \textit{Uniform\_IW} shows slightly higher MCD than the baseline 
because of the remaining audible noise generated from mu-law companding.

\begin{table}
\centering
\caption{GSNR (dB), SSNR (dB), and RMSE\(_{f0}\) (Hz) result. Higher is better for SNR and lower is better for RMSE\(_{f0}\)}\label{table:SNR}
\begin{tabular}{llll}
    \Xhline{3\arrayrulewidth}
    $\textbf{Methods}$ & $\textbf{GSNR}$ & $\textbf{SSNR}$ & $\textbf{RMSE}_{f0}$ \\ \hline
    Baseline\cite{kim2018flowavenet:7}    & -2.127 & -2.284  & 44.881  \\
    \textit{Uniform\_IW}    & -1.902 & -1.990 & 38.359  \\
    \textit{Gaussian\_Tanh} & -2.112 & -2.186 & 37.208   \\
    \textit{Flow\_Dense}   & -2.048 & -2.141  & 44.066   \\ 
    \Xhline{3\arrayrulewidth}
\end{tabular}
\end{table}

Table \ref{table:SNR} presents the result of SNR and RMSE\(_{f0}\). All proposed methods show higher SNR than baseline, indicating that audio dequantization can help reducing noise. In addition, \textit{Uniform\_IW} and \textit{Gaussian\_Tanh} dequantization show better performance in modeling fundamental frequency, while \textit{Flow\_Dense} dequantization shows a comparable result with the baseline model. We also visualized test outputs for qualitative evaluation in Figure \ref{fig:mel-spec}. Figure \ref{fig:mel-spec}(f) and Figure \ref{fig:mel-spec}(h) show clearer harmonic structures than Figure \ref{fig:mel-spec}(b). Although Figure \ref{fig:mel-spec}(d) shows less clear harmonic structures than other approaches, we can see that periodic noises are reduced significantly. We provide audio results on our online demo webpage.\footnote{\url{https://claudin92.github.io/deqflow_webdemo/}}

\section{Conclusions}
In this paper, we proposed various audio dequantization schemes that can be implemented in flow-based neural vocoder. For the uniform dequantization, we compressed the range of audio domain to match with conventional uniform dequantization method by using \textit{mu-law companding} compression. In addition, we implemented \textit{iw dequantization} to resolve the noise issue that occurs from the lossy compression. For the Gaussian dequantization, we applied hyperbolic tangent normalization on data-oriented Gaussian noise to properly fit the data within the audio range. Lastly, we modified flow block in flow-based neural vocoder to construct variational dequantization model to apply more flexible noise. From the experiments, we demonstrate that implementing audio dequantization can supplement the flow-based neural vocoder to produce better audio quality with fewer artifacts.

\clearpage
\bibliography{mypaper01}
\bibliographystyle{IEEEtran}
\end{document}